\PassOptionsToPackage{safe}{tipa}
\documentclass[cameraready]{Interspeech}
\title{\textsc{ReSUM}: Reasoning-Driven Multi-Agent Framework for Query-Focused Summarization}
\author[affiliation={1}]{Sonali}{}
\author[affiliation={2}]{Deeksha}{Varshney}
\author[affiliation={1}]{Asif}{Ekbal}
\address{$^1$Indian Institute of Technology Patna, India\\
         $^2$Indian Institute of Technology Jodhpur, India}
\email{sonali\_pa2521phd92@iitp.ac.in, deeksha@iitj.ac.in, asif@iitp.ac.in}

\usepackage{comment}
\usepackage{tcolorbox}

\begin{document}

\maketitle
\begin{abstract}
   LLMs excel at language tasks, but long-meeting QFS remains difficult due to complex speaker dynamics, sparse key information, and limited data. Meetings involve disfluencies, long transcripts and complex discourse. QFS have high risk of information omission , hallucination, even for advanced models such as GPT-4-mini and Gemini.  Existing end-to-end methods compress retrieval, evidence extraction, and summary generation into a single pass, failing to handle the complex reasoning required for targeted meeting summaries. We propose \textsc{ReSUM}, a modular multi-LLM framework coordinated by a planner that orchestrates retrieval and summarization agents. By distilling teacher reasoning into query- and plan-conditioned retrievers and summarizers, \textsc{ReSUM} enhances long-context robustness, minimizes irrelevant exploration, and generates more faithful and comprehensive query-focused summaries. Experiments show consistent gains on query-focused meeting summarization benchmarks. We make the code publicly available \footnote{https://anonymous.4open.science/r/ReSUM-CD12/}.
\end{abstract}

\section{Introduction}

\begin{figure}[h]
    \centering
    \includegraphics[
        width=\linewidth,
        height=0.30\textheight,
        keepaspectratio
    ]{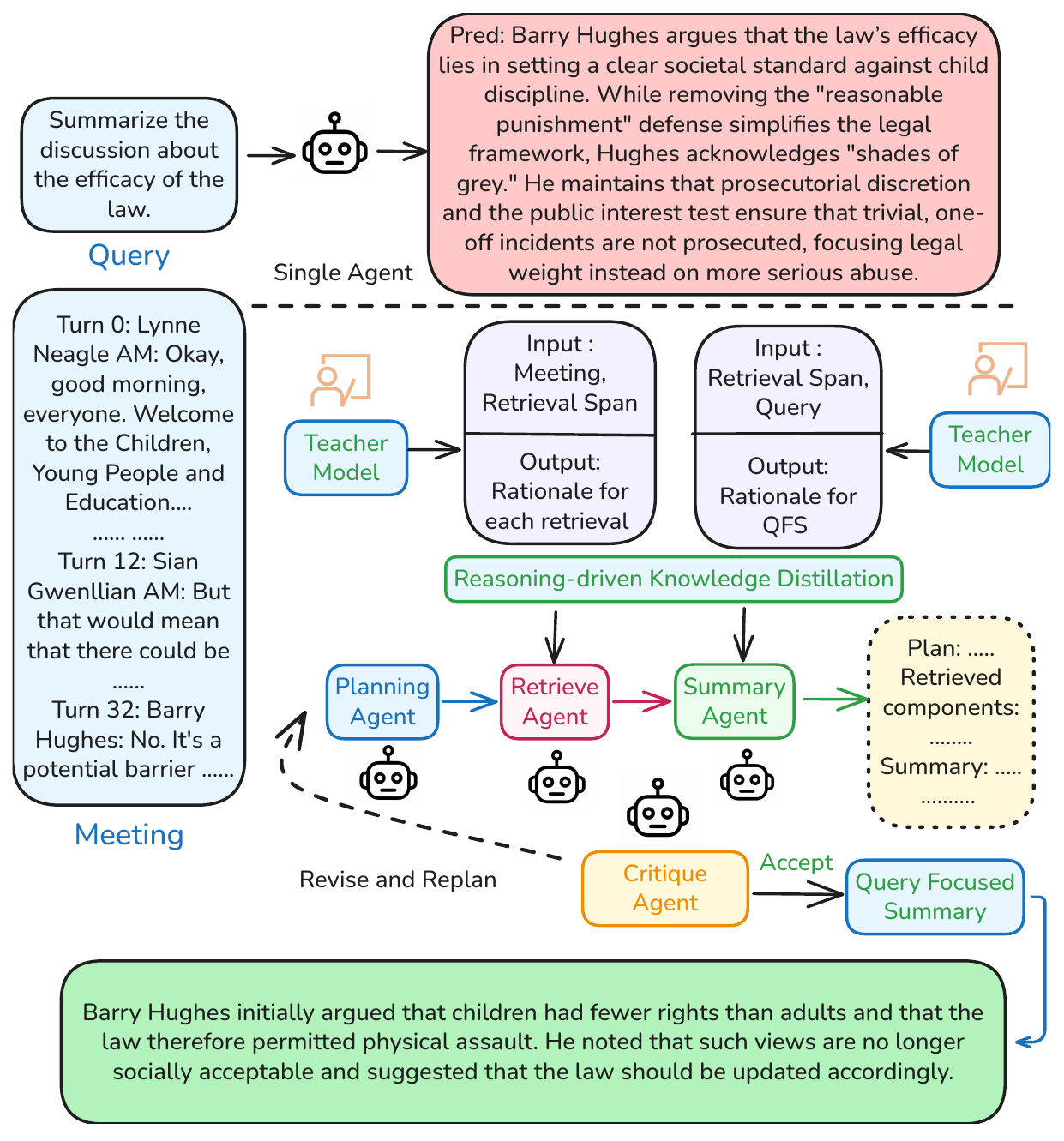}
    \caption{Single Agent that fails to capture query-specific context, our Multi-Agent pipeline divides tasks between specialized agents. By utilizing reasoning-driven knowledge distillation, the planning and retrieval agents ensure precise content selection, while the critique loop provides feedback-driven refinement for a highly accurate, query-focused summary.}
    \label{fig:introduction}
\end{figure}

Meetings and conversations are a common way for people to collaborate across different fields such as business, education, and research. These multi-speaker interactions often involve sharing ideas, discussing plans, and making decisions. To preserve the information discussed, they are frequently recorded and transcribed so that participants can review them later for reference. Unlike structured documents, meeting transcripts pose significant challenges due to interactive discourse, topic shifts, and the dispersion of relevant information across long sequences of utterances \cite{ijcai2022p764,kumar2022meetingsummarizationsurveystate}. These challenges are further exacerbated in query-focused scenarios, where the objective is to produce concise summaries tailored to specific information needs rather than broad, general overviews \cite{zhong2021qmsumnewbenchmarkquerybased}. In long meetings, query-relevant content is frequently scattered and interleaved with substantial irrelevant material, making effective evidence identification particularly challenging \cite{jiachen2024investigatingconsistencyquerybasedmeeting}.

Recent advances in LLMs have significantly improved abstractive summarization \cite{SUPRIYONO2024100070}, yet applying them to query-focused meeting summarization remains challenging. Fine-tuning LLM is computationally costly, annotated data is scarce, and the task requires jointly solving multiple subtasks: identifying query-relevant segments in long transcripts, extracting fine-grained evidence, and generating a faithful, coherent summary. Due to the length and structural complexity of meetings, single agent may miss salient information, motivating iterative retrieval and refinement. Moreover, effective query-focused meeting summarization demands multi-step reasoning over long meeting, including reasoning about relevance, redundancy, and evidence sufficiency capabilities that remain limited in smaller language models.

To address these limitations, we propose \textsc{ReSUM}, a modular multi-agent framework for query-focused meeting summarization that decouples long-context reasoning from text generation by distributing planning, retrieval, summarization, and critique across specialized agents A Planner agent performs long-context reasoning to construct an explicit traversal plan, identifying query-relevant subtopics and discourse regions and reducing unguided exploration of long transcripts. Guided by this plan, a rationale-distilled Retriever internalizes high-level reasoning during training to select evidence spans that are both topically aligned and query-focused. Summarizer then generates concise, coherent, and faithful summaries grounded exclusively in the retrieved evidence, while a Critique agent evaluates reasoning consistency, evidence sufficiency, summary factuality, and plan adherence, enabling iterative re-planning and refinement.

We adopt a reasoning-distilled training paradigm based on chain-of-thought (COT) supervision, where a strong teacher LLM produces explicit reasoning traces to justify retrieval decisions and evidence integration for summarization. These traces are used during supervised fine-tuning to train smaller LLM's to imitate the underlying reasoning along with generation. This distillation is applied consistently across agents, including both the Retriever and the Summarizer, enabling reasoning behaviors to be internalized not only for text generation but also for evidence selection and refinement. By coupling multi-agent decomposition with reasoning distillation, each agent learns both its functional role and the implicit reasoning required for robust coordination.

Our contributions are \textit{three-fold}:
\begin{enumerate}
\item We introduce \textsc{\textsc{ReSUM}}, a modular multi-agent framework for QFMS that decomposes long-context reasoning into planning, retrieval, critique, and synthesis, enabling iterative evidence accumulation over long meeting transcripts.

\item We propose a reasoning-aware training strategy via CoT distillation, where a strong teacher LLM supervises intermediate agent reasoning, transferring planning and reasoning abilities to smaller models.

\item We evaluate \textsc{ReSUM} on meeting transcripts (QMSum) and long-form narratives (SQuALITY), demonstrating consistent gains on complex and implicit queries across zero-shot, few-shot, fine-tuning, and RAG settings.

\end{enumerate}

\section{Related Work}

\textbf{Query focused Meeting Summarization (QFMS)}: QFMS generates summaries based on specific user queries, a task complicated by sparse, distributed information in long transcripts \cite{liu-xu-2023-learning}. To address this,  QMSum benchmark \cite{zhong2021qmsumnewbenchmarkquerybased} was introduced, favoring retrieve-then-summarize architectures. Early methods utilized hierarchical encoders and query-aware attention \cite{zhu-etal-2020-hierarchical}, while subsequent research explored coarse-to-fine summarization \cite{zhang-etal-2022-summn} and utterance ranking \cite{liu-xu-2023-learning}. Recent advancements leverage LLMs through Socratic pretraining \cite{pagnoni-etal-2023-socratic}, knowledge extraction \cite{yu-etal-2023-improving-query}, and multi-query prompting \cite{laskar-etal-2024-query}. However, most current systems rely on static, single-pass pipelines and lack mechanisms for iterative refinement.

\textbf{Reasoning in Agents: }

Recent work enhances language model reasoning through both prompting and training-based methods. CoT prompting elicits intermediate rationales and improves complex reasoning, with robustness further boosted by decoding strategies such as self-consistency and least-to-most prompting \cite{zhou2023leasttomostpromptingenablescomplex,10.5555/3600270.3601883,wang2023selfconsistencyimproveschainthought}. Structured prompting paradigms like Self-Ask explicitly decompose problems into subquestions \cite{press-etal-2023-measuring}. Beyond prompting, explicit CoT supervision using large-scale rationale datasets improves zero- and few-shot performance \cite{kim-etal-2023-cot}, and has been extended to logical deduction, symbolic planning, and teacher-student distillation \cite{liu-etal-2023-logicot,verma2025teachingllmsplanlogical,li-etal-2023-symbolic}. Reinforcement learning and self-critique further refine reasoning behaviors \cite{byun-etal-2024-ares}. More recent agentic frameworks embed reasoning within planning-oriented loops, interleaving reasoning and action or explicitly separating planning from execution and exploring multiple reasoning paths \cite{yao2023reactsynergizingreasoningacting,wang-etal-2023-plan,10.5555/3666122.3666639}.

\textbf{Multi-agent system: }Recent research explores multi-agent LLM systems that collaborate through structured natural-language interactions. General frameworks like AutoGen \cite{wu2023autogenenablingnextgenllm} and role-based prompting in CAMEL \cite{li2023camelcommunicativeagentsmind} facilitate agent coordination, while AgentVerse \cite{ICLR2024_578e65cd} demonstrates that specialized agent teams outperform single models in reasoning and coding.  Performance relies on explicit planning and deliberation: multi-agent debate enhances factuality \cite{du2023improvingfactualityreasoninglanguage}, and recent benchmarks highlight the importance of communication design \cite{zhu-etal-2025-multiagentbench}. To support iterative improvement, systems now incorporate long-term memory \cite{10.1145/3586183.3606763} and verbal self-reflection mechanisms like Reflexion \cite{shinn2023reflexionlanguageagentsverbal} to refine behavior without retraining.

\section{Methodology}

\begin{figure*}[!h]
    \centering
    \includegraphics[width=\linewidth]{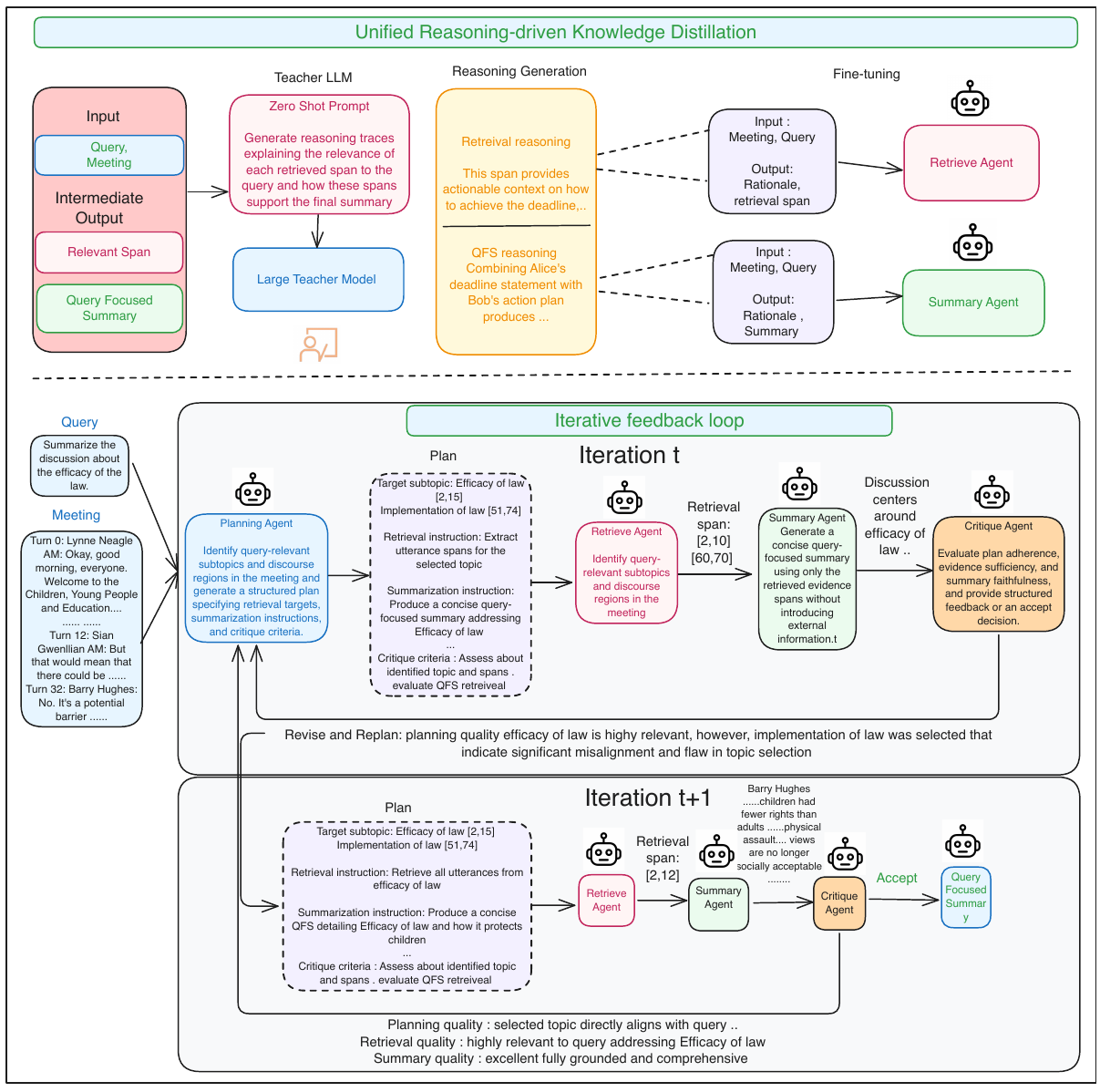}
    \caption{Architectural Overview of \textsc{ReSUM}, where the upper part illustrates the training stage with teacher–student distillation and reasoning supervision for the Retriever and Summarizer, and the lower part depicts the multi-agent pipeline coordinating Planner, Retriever, Summarizer, and Critique agents for iterative, query-focused meeting summarization.}
    \label{fig:methodology}
\end{figure*}

\textbf{Task Formulation: }Given a query $Q$ and a meeting transcript 
\begin{equation}
M = \{ u_t \mid t = 1, \ldots, T \}
\end{equation}, query-focused meeting summarization aims to generate a concise and faithful summary $S$ conditioned on $Q$ using only relevant evidence. To handle long, multi-speaker transcripts with scattered information, we used a collaborative LLM multi-agent system with descriptive prompts and reasoning-based fine-tuning.  The system consists of four agents: (1) the Planner agent, (2) the Retriever agent, (3) the Summarizer agent, and (4) the Critique agent, \begin{equation}
\mathcal{L} = {L_p, L_r, L_s, L_c}
\end{equation} which collectively decompose the task into planning, evidence selection, grounded summarization, and iterative validation.
\newline
Given $(Q, M)$, the Planner agent $L_p$ generates a traversal plan that identifies query-relevant subtopics and discourse regions, constraining the search space. Guided by this plan, the Retriever agent $L_r$ selects query-relevant turns evidence from $M$. The Summarizer $L_s$ then generates a query-focused summary grounded exclusively in the retrieved evidence. A Critique agent $L_c$ evaluates plan adherence, evidence sufficiency, and summary faithfulness, providing feedback to these agents for iterative refinement. Overall, this defines a mapping 
\begin{equation}
S = \mathcal{F}_{\mathcal{L}}(Q, M)
\end{equation}
that supports selective reasoning over long meeting transcripts.

\subsection{\textsc{ReSUM} Components}

\subsubsection{Planning Agent}
The Planning Agent initiates QFS by decomposing the original task into series of steps for long meeting transcripts. The planner is initiated with prompt, conditioned on the user query $Q$, meeting transcript $M$, agent descriptions $D$ as shown in Table ~\ref{fig:planner}. It comes up with step-by-step plan and delegates sub-tasks to downstream agents. 

\begin{table}[ht]
\caption{Planning Agent: Role definition and multi-agent coordination instructions.}
    \label{fig:planner}
\centering
\scriptsize
\renewcommand{\arraystretch}{1.4}
\begin{tabular}{|p{7.5cm}|}
\hline
\textbf{Instructions} \\
\hline
You are a Planner agent for long meeting QFS. Given user query Q, meeting transcript M, and current state S, construct a structured, dependency-aware execution plan for agents. 

\vspace{0.5em}
Given agents description: Retriever (select fine-grained, query-relevant utterance spans from planner-selected discourse regions, focusing on evidence-bearing content), Summarizer (generate a query-focused summary grounded strictly in retrieved spans without hallucination), and Critique (evaluate planning quality, retrieval relevance, and summary faithfulness).

\vspace{0.5em}
Generate a step-by-step plan for each agent, to generate a query-focused summary. \\
\hline
\end{tabular}

\end{table}

Under this formulation, the Planner constructs a global execution plan for all agents. Given the query $Q$ and meeting transcript \begin{equation}
M = \{u_t \mid t = 1,\ldots,T\}\end{equation}, it generates an explicit plan that specifies (i) the subtopics or discourse regions to explore, (ii) instructions for generating query-relevant utterance spans to be retrieved, (iii) instructions for producing a summary grounded in the retrieved evidence, and (iv) evaluation criteria for the Critique agent. After each iteration, each agent output with final feedback of critique agent is returned to the Planner, using which the Planner continues to revise its plan and delegate further task. This planning process constrains long-context exploration and enables targeted QFS.

At planning iteration $i$, the Planner conditions on the query $Q$, meeting transcript $M$, a task prompt $P_T$, and a structured state $\mathcal{S}^{(i-1)} $that captures retrieved evidence, intermediate summaries, and critique feedback.  The Planner input is defined as \begin{equation}\mathbf{x}^{(i)} = \bigl[ Q, M, \mathcal{S}^{(i-1)}, P_T \bigr]\end{equation}. This input is provided to the Planner agent $L_p$, which produces a structured multi-agent execution plan: \begin{equation}Plan^{(i)} = L_p\!\left( \ell\!\left( \mathbf{x}^{(i)} \right) \right)\end{equation}, where $\ell(\cdot)$ denotes planner prompt. After this,  Planner starts by delegating these tasks to agents. After each iteration, Iterative feedback by critique agent is incorporated for subsequent planning iterations, allowing the Planner to refine traversal decisions and sub-task assignments. Figure ~\ref{fig:methodology} shows an example of the Planner generating the plan for each agent iteratively. We limit the maximum number of planning iterations to 3 to reduce computational cost.

\subsubsection{Query-Focused Retriever with Reasoning Supervision}

Unlike traditional chunk-retrieve-rank pipelines that rely on static indexes and similarity-based ranking, we propose a query-focused, reasoning-guided Retriever tailored for long and structured meeting transcripts. Instead of retrieving from entire transcript, the Retriever operates under explicit guidance from the Planner and restricts its search to planner-selected topic segments. After the Planner generates a complete execution plan at iteration $i$, denoted $plan^{(i)}$, the Retriever operates over the subset of discourse units $\mathcal{M}^{(i)} \subset \mathcal{M}$ specified in that plan. Conditioned on the user query $Q$ and the planner-defined subset $\mathcal{M}^{(i)}$, the Retriever selects relevant span
\begin{equation}
\mathcal{R}_\theta(Q, plan^{(i)}, \mathcal{M}^{(i)}) \rightarrow \{s_k\},
\end{equation}
where each $s_k$ is a fine-grained span aligned with discourse boundaries.

\begin{table}[ht]
\caption{ Retriever Agent: System prompt and generated rationale for evidence extraction.}
\label{fig:retre}
\centering
\scriptsize
\renewcommand{\arraystretch}{1.4}
\begin{tabular}{|p{7.5cm}|}
\hline
\textbf{Prompt} \\
\hline
Given (i) a meeting transcript with turn-level identifiers, (ii) a user query, and (iii) a relevant span defined by its start and end turn IDs along with the corresponding span text, generate a structured rationale explaining how the span addresses the query. 

\vspace{0.5em}
The rationale should: (1) identify key evidence lines within the span, (2) explain their relevance to the query, (3) articulate any implicit reasoning linking the span to the query, and (4) rely exclusively on transcript content within the specified turn range.

\vspace{0.5em}
The output must present a coherent, step-by-step explanation grounded in the provided span. \\
\hline
\textbf{Rationale} \\
\hline
The Project Manager introduces the prototype by outlining the meeting agenda and emphasizing trend-driven design elements, such as a fashionable and fruit-inspired look. The User Interface then describes the prototype's banana-shaped design, directly aligning with these trends. Together, these details explain how they introduced the prototype in response to the query. \\
\hline
\end{tabular}

\end{table}

We adopt a teacher-student distillation framework. A large teacher LLM (e.g., GPT-4) generates reasoning-based rationales explaining why each gold span from meeting $s_k^\ast$ is relevant for a given query. As illustrated in Table ~\ref{fig:retre}, given a dataset with source meeting , query and relevant span we generated a rationale $r_k$ that explains how the given span (defined by its turn IDs and text) answers the user query for each input. It also identifies key evidence phrases, explain their relevance, and clarify any implicit reasoning connecting the span to the query. We incorporated rationales into the training process of smaller LLM.

The  Retriever is fine-tuned using a distillation objective for supervise span selection and rationale alignment:

\begin{equation}
\mathcal{L}_{\text{distill}} 
= \sum_k \mathrm{CE}\big(\mathcal{R}_\theta(Q, \mathcal{M}^{(i)}), (s_k^*, r_k)\big)
\end{equation}
, which allows it to internalize reasoning patterns without requiring explicit COT generation at inference. This enables the Retriever to implicitly encode reasoning-informed relevance, improving retrieval of semantically and contextually appropriate spans. The rationale generation loss enables the model to learn to generate the intermediate reasoning steps for the prediction, and could therefore guide the model in better predicting the relevant span for the query. Retrieved spans are stored in a structured state $S$ and passed to the Summarizer for evidence-grounded, query-aligned summary generation, tightly coupling query intent, planning, and retrieval. We prepend task prefixes (\texttt{[relevant\_span]}, \texttt{[rationale]}) to the input examples and train the smaller model to output $\hat{s}_i$ when \texttt{[relevant\_span]} is provided, and to produce $\hat{r}_i$ when \texttt{[rationale]} is given.

\subsubsection{Query-Focused Summarizer with Reasoning Traces}

Once the relevant span for query is generated using Retrieval agent, the Summarizer produces a query-focused summary on the meeting span content selected by the Retriever. Instead of attending to the full transcript, the Summarizer is conditioned solely on the retrieved spans $\{s_k\}$, ensuring that the generated summary is both relevant and factually grounded. Formally, given a user query $Q$ and retrieved spans \begin{equation}\{s_k\} = \mathcal{R}_\theta(Q, \mathcal{M}^{(i)})\end{equation}, the Summarizer outputs \begin{equation} QFS\_Summary^{(i)} = \mathcal{S}_\phi(Q, \{s_k\})\end{equation}, where $\mathcal{S}_\phi$ denotes the Summarizer model and $QFS\_Summary^{(i)}$ represents the query-focused summary at iteration $i$.

\begin{table}[ht]
    \caption{Summarizer Agent: Instructional prompt and evidence-grounded rationale.}
    \label{fig:summ}
\centering
\scriptsize
\renewcommand{\arraystretch}{1.2}
\begin{tabular}{|p{7.5cm}|}
\hline
\textbf{Prompt} \\
\hline
Given (i) a user query, (ii) a relevant transcript span, and (iii) a query-focused summary, generate a structured rationale explaining how the span supports the summary with respect to the query. 

\vspace{0.5em}
The rationale should: (1) Identify key evidence lines from the span, (2) Explain how each summary point addresses the query, (3) Clarify implicit reasoning linking the span to the summary, and (4) Rely only on information within the provided span.

\vspace{0.5em}
Produce a clear, step-by-step explanation grounded strictly in the given content. \\
\hline
\textbf{Rationale} \\
\hline
The team introduced a fashion-inspired prototype, described as bright yellow and banana-shaped. They emphasized simplicity, ergonomics, and usability for both left- and right-handed users, along with advanced technology. The discussion about size and design choices supports the summary's focus on creativity, functionality, and ease of use. \\
\hline
\end{tabular}

\end{table}

We generated intermediate reasoning traces from teacher LLM reasoning traces $r_k^{(\text{sum})}$ that explains 
 how each retrieved span contributes to the final summary. As illustrated in Table ~\ref{fig:summ}, Given a query and its corresponding relevant span, we generate a rationale 
$r_k$ that explains how the span answers the query by highlighting key evidence phrases, clarifying their relevance to each summary point, and making implicit reasoning steps explicit.

These rationales are incorporated into the training process of smaller LLMs to improve their ability to produce query-grounded, evidence-aligned summaries. The Summarizer agent is fine-tuned to generate query-focused summary with rationale, analogous to the retrieval agent. 
The Summarizer is trained to jointly generate the query-focused summary and its corresponding rationale via a distillation objective:
\begin{equation}
\mathcal{L}_{\text{sum}} 
= \sum_k \mathrm{CE}\Big(
\mathcal{S}_\phi(Q, \{s_k\}), 
[\, y_k^\ast \; ; \; r_k^\ast \,]
\Big),
\end{equation}
where $y_k^\ast$ denotes the query-focused summary and $r_k^\ast$ denotes the corresponding teacher-generated rationale explaining how span $s_k$ answers the query.

Integrating reasoning traces enables the model to generate the summary while explicitly aligning outputs with retrieved spans to reduce hallucination. By conditioning solely on evidence-relevant content, the model avoids irrelevant context,omission and hallucination. These intermediate traces guide the Summarizer to understand why each span matters, producing summaries that are faithful, query-focused, and grounded in retrieved evidence. The resulting summary is stored in the structured execution state \begin{equation}\mathcal{S}^{(i)} \leftarrow QFS\_Summary^{(i)}\end{equation}, making it accessible for subsequent Critique, Refinement, or multi-agent iterative processing.

\subsubsection{Iterative Critique Framework}
The Critique agent (Table \ref{fig:critic}) acts as the evaluation and control module, systematically verifying planning, retrieval, and summarization agents outputs for correctness, coherence, and alignment with Planner objectives. This ensures that the multi-agent system avoids ``self-confirmation bias," where a generator might overlook its own logical inconsistencies.

\begin{table}[ht]
    \caption{Critique Agent: Evaluation and refinement instructions.}
    \label{fig:critic}
\centering
\scriptsize
\renewcommand{\arraystretch}{1.4}
\begin{tabular}{|p{7cm}|}
\hline
\textbf{Instructions} \\
\hline
You are a Critique agent in a multi-agent QFS system. Evaluate outputs from the Planner, Retriever, and Summarizer, and provide actionable feedback.

\vspace{0.5em}
\textbf{1. Planner (Planning Alignment)} \\
(i) Are selected topics appropriate for the query and meeting context? 
(ii) Were necessary agents correctly invoked and assigned roles? 
(iii) Are instructions clear and query-aligned?

\vspace{0.5em}
\textbf{2. Retriever (Retrieval Quality)} \\
(i) Are retrieved turn-ID ranges directly relevant to the query? 
(ii) Are retrieved spans evidence-bearing rather than loosely related? 
(iii) Is important query-relevant content missing?

\vspace{0.5em}
\textbf{3. Summarizer (Summary Quality)} \\
(i) Is the summary grounded only in retrieved evidence? 
(ii) Does it fully and correctly answer the query? 
(iii) Is it faithful, coherent, and free of hallucinations?

\vspace{0.5em}
Conclude whether the overall output is acceptable or requires refinement. \\
\hline
\end{tabular}

\end{table}

\begin{itemize}
    \item Retriever outputs: Check if retrieved spans $\{s_k\}$ are query-relevant, topic-aligned, and coherent with Planner-selected segments $\mathcal{M}^{(i)}$. It explicitly checks for contextual entailment, ensuring the retrieved evidence provides sufficient logical support for the given query.
    \item Summarizer outputs: Assess $QFS\_Summary^{(i)}$ for faithfulness, evidence grounding, and coverage of essential content.
\end{itemize}
 Critique prompts incorporate evaluation criteria $P_V$, the query $Q$, relevant spans, summaries, and past state: \begin{equation}{prompt}^{(i)} = \big\{ P_V, Q, \text{plan}^{(i)}, \text{span}^{(i)}, QFS^{(i)}, \mathcal{S}^{(i)} \big\}\end{equation}.
It produces a structured feedback report: \begin{equation}\mathcal{F}^{(i)} = \text{Critique}\big(\text{plan}^{(i)}, \{s_k\}, QFS^{(i)}\big)\end{equation}
stored in the execution state $\mathcal{S}^{(i)}$ and fed back to the Planner for iterative refinement until an \texttt{accept} signal or iteration limit (3) is reached. It serves as a corrective signal for the next execution state.

To ensure robust coordination, the system employs a stateful refinement cycle where the execution state $S^{(i)}$ retains prior reasoning trajectories and failures. This persistent memory prevents convergence to local optima or oscillation between inconsistent states. Termination is governed by a formal acceptance criterion requiring all Boolean constraints in the evaluation rubric $P_V$ to be satisfied. To control computational cost, a strict iteration limit enforces a reflection budget.

\section{Datasets and Experiments}
We evaluate our framework on two benchmarks: QMSum \cite{zhong2021qmsumnewbenchmarkquerybased}, a multi-domain meeting dataset requiring evidence extraction from noisy, multi-speaker transcripts to generate query-focused summary; and SQuALITY \cite{wang-etal-2022-squality}, which demands long-context reasoning over narrative short stories with multi-reference evaluation.

We compare our framework against two types of long-context summarization baselines- (1) End-to-End Generation: models that summarize the entire transcript at once, tested with both zero-shot and CoT prompting, and (2) RAG:  using external sparse and dense retrievers to select the top-$K$ relevant chunks for summarization. We evaluate both reasoning and non-reasoning LLM's across open and closed-source settings. Non-reasoning models include LLaMA-3.2~3B and Gemma-3~4B, while reasoning-based models  Qwen-3~4B and DeepSeek-R1-Distill-Qwen-1.5B were tested. Closed-source models, GPT-4-mini are also included. All baselines follow their original configurations for fair comparison. In our modular setup, the planner and critique use GPT-4-mini, the retriever employs Qwen-3~4B and DeepSeek-R1-Distill-Qwen-1.5B, and the summarizer relies on LLaMA-3.2~2B and Gemma-3~4B.
\textbf{Evaluation Metrics}
We evaluate retrieval using standard metrics: \textsc{Recall@k}, \textsc{Precision@k}, \textsc{MAP@k}, \textsc{Hit@k}, \textsc{MRR@k}, \textsc{F1@k}, and \textsc{Overlap@k}. Summaries are assessed with automatic metrics such as \textsc{ROUGE-1/2/L} (R1,R2,RL) \cite{lin-2004-rouge}, \textsc{BERTScore} (BS) \cite{zhang2020bertscoreevaluatingtextgeneration}, \textsc{MoverScore} \cite{zhao-etal-2019-moverscore}, \textsc{COMET} \cite{rei-etal-2020-comet}, and \textsc{chrF} \cite{popovic-2015-chrf}.

\section{Results and Analysis}

\begin{table*}[ht]
\caption{Evaluation results on QMSum and SQuALITY.}
\label{quantresult}
\centering
\scriptsize
\renewcommand{\arraystretch}{1.3} 
\begin{tabular}{|l|ccccccc|ccccccc|}
\hline
\textbf{Model} & \multicolumn{7}{c|}{\textbf{QMSum}} & \multicolumn{7}{c|}{\textbf{SQuALITY}} \\
\hline
& R1 & R2 & RL & BS & Mover & COMET & chrF & R1 & R2 & RL & BS & Mover & COMET & chrF \\
\hline
\multicolumn{15}{|c|}{\textit{Zero-shot}} \\
\hline
LLaMA3-3B      & 0.27 & 0.06 & 0.17 & 0.15 & 0.53 & 0.66 & 34.01 & 0.32 & 0.06 & 0.17 & 0.08 & 0.53 & 0.67 & 35.52 \\
Gemma3-4B      & 0.20 & 0.04 & 0.13 & 0.06 & 0.52 & 0.65 & 30.39 & 0.32 & 0.06 & 0.17 & 0.04 & 0.53 & 0.66 & 37.38 \\
DeepSeek-1.5B  & 0.17 & 0.02 & 0.13 & 0.00 & 0.51 & 0.57 & 13.10 & 0.26 & 0.04 & 0.15 & 0.04 & 0.52 & 0.59 & 31.99 \\
Qwen3-4B       & 0.27 & 0.07 & 0.17 & 0.16 & 0.54 & 0.68 & 35.22 & 0.32 & 0.07 & 0.18 & 0.11 & 0.53 & 0.67 & 32.35 \\
GPT-4-mini     & 0.31 & 0.09 & 0.20 & 0.18 & 0.55 & 0.71 & 37.78 & 0.35 & 0.08 & 0.19 & 0.13 & 0.54 & 0.72 & 39.34 \\
\hline
\multicolumn{15}{|c|}{\textit{Chain-of-Thought (CoT)}} \\
\hline
COT LLaMA      & 0.28 & 0.07 & 0.18 & 0.14 & 0.53 & 0.66 & 35.03 & 0.33 & 0.07 & 0.18 & 0.10 & 0.53 & 0.64 & 32.56 \\
COT Qwen       & 0.28 & 0.08 & 0.18 & 0.17 & 0.54 & 0.68 & 35.39 & 0.32 & 0.07 & 0.18 & 0.12 & 0.54 & 0.67 & 34.49 \\
\hline
\multicolumn{15}{|c|}{\textit{Retrieval-Augmented Generation (RAG)}} \\
\hline
RAG-FAISS      & 0.28 & 0.08 & 0.19 & 0.21 & 0.54 & 0.57 & 25.48 & 0.14 & 0.03 & 0.10 & 0.03 & 0.51 & 0.52 & 07.91 \\
RAG-BM25       & 0.30 & 0.08 & 0.20 & 0.21 & 0.54 & 0.58 & 27.18 & 0.16 & 0.04 & 0.11 & 0.03 & 0.51 & 0.49 & 09.18 \\
RAG-HYBRID     & 0.29 & 0.08 & 0.20 & 0.22 & 0.54 & 0.63 & 28.81 & 0.17 & 0.04 & 0.12 & 0.05 & 0.52 & 0.50 & 08.39 \\
RAG-CONTRIEVER & 0.30 & 0.08 & 0.20 & 0.23 & 0.54 & 0.65 & 29.02 & 0.16 & 0.04 & 0.11 & 0.05 & 0.52 & 0.54 & 08.45 \\
\hline
\multicolumn{15}{|c|}{\textit{Fine-tuned (FT)}} \\
\hline
FT LLaMA       & 0.29 & 0.09 & 0.20 & 0.18 & 0.56 & 0.69 & 38.62 & 0.35 & 0.09 & 0.19 & 0.09 & 0.54 & 0.65 & 26.84 \\
FT Qwen        & 0.33 & 0.11 & 0.24 & 0.22 & 0.60 & 0.71 & 38.75 & 0.36 & 0.09 & 0.20 & 0.12 & 0.55 & 0.69 & 34.63 \\
\hline
\multicolumn{15}{|c|}{\textit{End-to-End Fine-tuned with Reasoning}} \\
\hline
FT+Res LLaMA   & 0.28 & 0.07 & 0.19 & 0.20 & 0.53 & 0.61 & 24.85 & 0.38 & 0.09 & 0.20 & 0.11 & 0.54 & 0.65 & 35.74 \\
FT+Res Qwen    & 0.35 & 0.12 & 0.28 & 0.22 & 0.60 & 0.75 & 39.50 & 0.40 & 0.14 & 0.20 & 0.19 & 0.61 & 0.65 & 38.00 \\
\hline
\multicolumn{15}{|c|}{\textit{\textsc{ReSUM} Multi-Agent}} \\
\hline
\textbf{Our Method} & \textbf{0.38} & \textbf{0.13} & \textbf{0.33} & \textbf{0.25} & \textbf{0.63} & \textbf{0.77} & \textbf{41.65} & \textbf{0.44} & \textbf{0.13} & \textbf{0.23} & \textbf{0.19} & \textbf{0.70} & 0.62 & \textbf{43.78} \\
\hline
\end{tabular}

\end{table*}

\subsection{Quantitative Results for Query Focused Summarization}
Table~\ref{quantresult} presents automatic evaluation results on QMSum and SQuALITY across zero-shot, chain-of-thought, retrieval-augmented, fine-tuned, and fine-tuning with rationales. The proposed ReSUM framework achieves the strongest overall performance on both datasets, improving consistently across lexical and semantic metrics. On QMSum, ReSUM yields higher ROUGE-1, ROUGE-2, and ROUGE-L scores than all baselines, along with gains in BERTScore, MoverScore, and chrF, indicating improved coverage, relevance, and semantic alignment in query-focused meeting summaries.

On SQuALITY, ReSUM similarly demonstrates robust performance, outperforming competing approaches on ROUGE-based metrics while maintaining strong semantic similarity scores. While retrieval-augmented methods show reasonable performance on QMSum, their effectiveness degrades on SQuALITY, suggesting limitations of static retrieval under longer and more diverse discourse structures. In contrast, ReSUM maintains stable improvements across both datasets, highlighting the benefits of explicit planning, reasoning-guided evidence selection, and iterative critique for long-context QFS.

\subsection{Ablation Studies}

\begin{table}[h]
\caption{Ablation study of core agents in \textsc{ReSUM} on QMSum using ROUGE-L.}
\label{tab:ReSUM-ablation}
\centering
\small
\begin{tabular}{p{5.0cm} c}
\toprule
Method & ROUGE-L \\
\midrule
\textsc{ReSUM} & 0.38 \\
w/o Planner & 0.32 ($\downarrow$0.06) \\
w/o Reasoning Retriever & 0.31 ($\downarrow$0.07) \\
w/o Critique & 0.33 ($\downarrow$0.05) \\
\bottomrule
\end{tabular}

\end{table}

We conduct ablation experiments on each agent of \textsc{ReSUM}, with results reported in Table~\ref{tab:ReSUM-ablation}. Removing the reasoning-guided Retriever leads to the largest performance degradation, with a drop of 0.636 ROUGE-L, highlighting its critical role in identifying query-relevant evidence. Eliminating the Planner also results in a substantial decline of 5.06 ROUGE-L, highlighting the importance of structured traversal and targeted exploration over long meeting transcripts. Removing the Critique agent causes a smaller yet notable reduction of 4.66 ROUGE-L, reflecting its contribution to iterative refinement and error correction.

\textbf{Importance of Planning Agent:} The Planning agent is central to \textsc{\textsc{ReSUM}}, as it jointly reasons over the user query, meeting subtopics, agent roles, and critique feedback to transform QFS from unguided retrieval into structured, dependency-aware multi-agent execution. By constraining exploration to query-relevant discourse regions and assigning precise sub-tasks to the Retriever, Summarizer, and Critique agents, the planner enables effective evidence selection and iterative refinement. As shown in Table~\ref{tab:ReSUM-ablation}, removing the planner forces retrieval over the full transcript, substantially expanding the search space and increasing the number of retrieved spans, which lowers retrieval precision due to topically misaligned evidence. This noise propagates downstream, causing the Summarizer to condition on irrelevant content, leading to reduced faithfulness, weaker query alignment, and increased hallucination.

\textbf{Importance of Retrieval Agent:}
Removing the Retriever agent degrades evidence selection and abstraction into the Summarizer, forcing it to implicitly scan the full transcript to identify query-relevant turns before summarization. This substantially expands the search space and increases retrieval errors with generation, leading to a sharp drop in retrieval precision and increased inclusion of  misaligned turns for the given query. The resulting noisy evidence degrades summary faithfulness and query alignment, while also increasing computational cost and hallucination risk due to long, weakly grounded contexts.

\textbf{Importance of Critique Agent:} The Critique agent functions as the verification and control loop of \textsc{ReSUM}, assessing retrieved spans and summaries against planner-defined criteria (relevance, coverage, faithfulness, redundancy) and generating structured feedback for refinement. By detecting missing evidence, over-retrieval, and misalignment, it prevents error propagation through targeted re-retrieval and summary correction. Without the Critique agent, the system becomes single-pass, leaving errors uncorrected and leading to coverage gaps, increased hallucinations, and lower accuracy.

\subsection{Substituting different LLMs}
\begin{figure}
\centering
\includegraphics[width=\columnwidth]{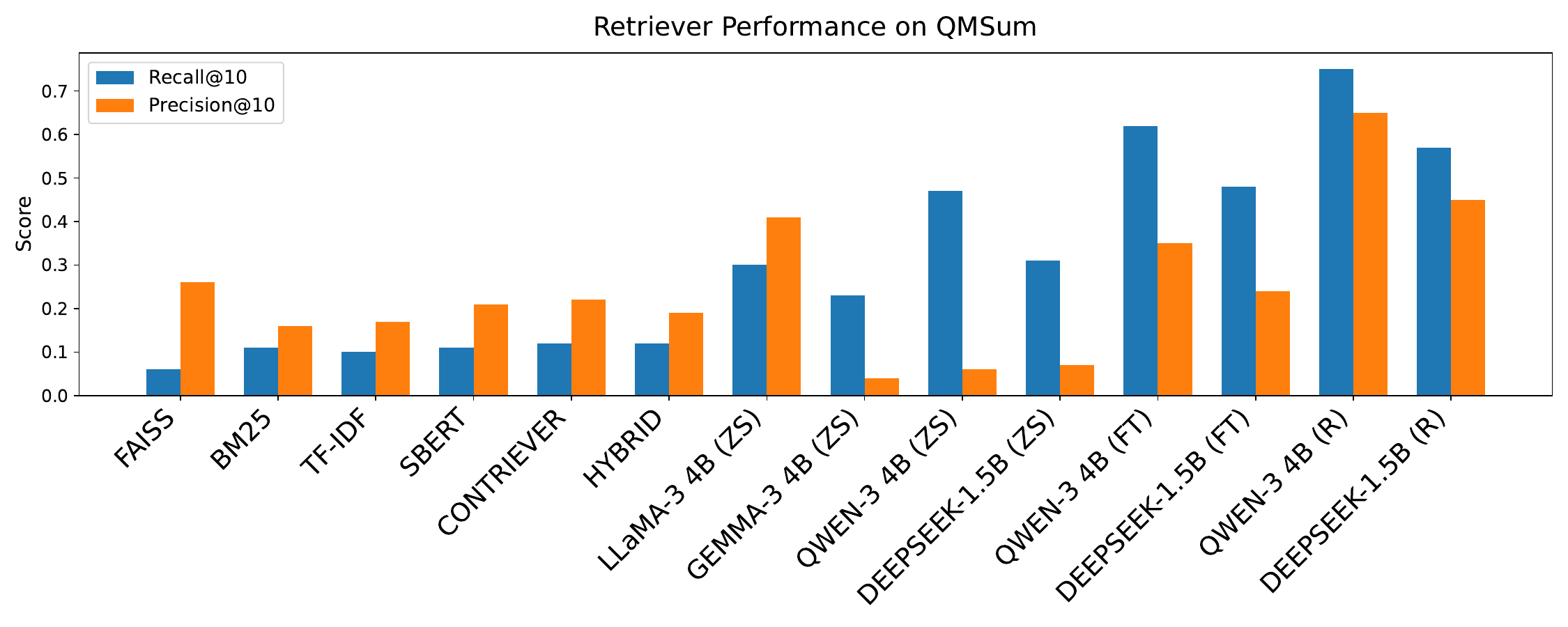}
\caption{Evaluating QMSum retriever performance via Recall@k and Precision@k: A comparative study of sparse, dense, hybrid, Zero-shot LLM (ZS), FT models, reasoning-driven fine-tuned (R) models.}
\label{fig:retriever_qmsum}
\end{figure}

\textbf{Different LLMs for Retriever:} Figure~\ref{fig:retriever_qmsum} compares different retrieval strategies on QMSum and shows a clear progression from sparse to reasoning-aware retrieval. Sparse retrievers such as BM25 and TF-IDF obtain low recall and precision due to their reliance on surface-level lexical overlap in long, paraphrase-rich meeting transcripts. Dense retrievers including SBERT and Contriever provide moderate improvements by capturing semantic similarity, while hybrid retrieval further increases recall but at the cost of reduced precision.
Zero-shot LLM-based retrievers show high variance: Qwen3-4B achieves higher recall but lower precision than LLaMA-3B. Supervised fine-tuning improves both metrics, with reasoning-driven supervision yielding the largest gains. Qwen3-4B under reasoning fine-tuning achieves the best recall–precision trade-off, reaching 0.7580 recall and 0.6480 precision, demonstrating that explicit reasoning supervision improves query-focused utterance selection while reducing irrelevant meeting turns.

\begin{figure}[t]
\centering
\includegraphics[width=\columnwidth]{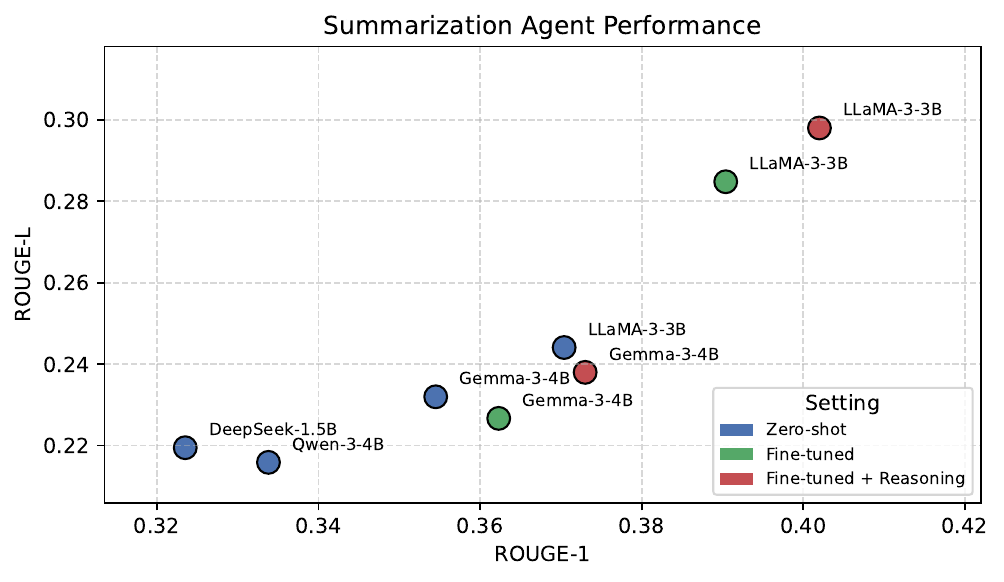}
\caption{Performance of summarization agents on QMSum across different settings. 
Zero-shot models, fine-tuned models, and reasoning-driven fine-tuned models. Each point represents a model.}
\label{fig:summariser_qmsum}
\end{figure}

\textbf{Different LLM Performance for Summarizer Agent:} Figure~\ref{fig:summariser_qmsum} compares zero-shot, fine-tuned, and reasoning-driven fine-tuned summarization performance on QMSum. In the zero-shot setting, LLaMA-3B achieves the strongest ROUGE-1 and ROUGE-L scores of 0.370 and 0.244, while Qwen-4B and DeepSeek-1.5B perform competitively on semantic metrics such as COMET and BERTScore. Fine-tuning improves performance across all models, with LLaMA-3B showing gains of 0.020 on ROUGE-L and 0.041 on ROUGE-2, indicating improved relevance and coverage. Reasoning-augmented fine-tuning yields further improvements, with LLaMA-3B reaching a ROUGE-1 score of 0.402 and a ROUGE-2 score of 0.172, and Gemma-3B achieving ROUGE-1 and ROUGE-2 scores of 0.373 and 0.138. The largest improvements are observed on semantic metrics, suggesting that explicit reasoning supervision enhances both lexical alignment and deeper semantic fidelity, particularly for larger models.

\subsection{Qualitative Analysis}

\textbf{Human Evaluation:} HHuman evaluation of the query-focused summaries is conducted by three linguistically trained annotators, who rate each summary on four criteria—relevance to the query, factual accuracy, fluency and readability, and completeness without unnecessary information—using a 1–5 Likert scale. To reduce bias, summaries are shuffled and accompanied by gold-span–based references from the source. Annotators are compensated according to university guidelines. Inter-rater reliability, measured using Cohen’s Kappa, yields a score of 0.72, indicating substantial agreement.

\begin{table}[ht]
\scriptsize
\centering
\caption{Human evaluation scores on a 1-5 Likert scale. Rel - Relevance, Fact - Factuality, Complete - Completeness}
\label{tab:human_eval}
\begin{tabular}{lccccc}
\toprule
\textbf{Model} & \textbf{Rel} & \textbf{Fact} & \textbf{Fluency} & \textbf{Complete} & \textbf{Mean} \\
\midrule
0-Shot & 1.1 & 1.0 & 3.5 & 1.0 & 1.65 \\
CoT & 1.4 & 1.2 & 3.2 & 1.2 & 1.75 \\
RAG & 3.5 & 3.4 & 2.8 & 3.2 & 3.23 \\
FT & 4.2 & 3.8 & \textbf{4.5} & 4.1 & 4.15 \\
\midrule
\textbf{Ours} & \textbf{4.5} & \textbf{4.5} & 4.3 & \textbf{4.4} & \textbf{4.43} \\
\bottomrule
\end{tabular}

\end{table}

\textbf{Case Studies: }
Table~\ref{tab:case_ex1} presents a multi-agent QFS example demonstrating how the Planner, Retriever, Summarizer, and Critique agents collaborate through iterative refinement. In Iteration 1, misaligned planning (requesting non-existent spans) propagates errors to retrieval and summarization, resulting in incomplete query coverage, which the Critique agent correctly flags as a structural inconsistency. In Iteration 2, the Planner restricts the scope to verifiable evidence, enabling precise retrieval and a summary that captures the legal nuance of the Public Interest Test, thereby improving evidence alignment, factual grounding, and overall response fidelity.

\begin{table}[ht]
\caption{Case Study 1: Summarize the efficacy of law. Initial outputs contain errors or omissions, which are corrected in subsequent iterations through refined planning and retrieval. }
\label{tab:case_ex1}
\centering
\scriptsize
\renewcommand{\arraystretch}{1.4}
\begin{tabular}{|l|p{5.7cm}|}
\hline
\multicolumn{2}{|c|}{\textbf{Iteration 1: Initial (Failure)}} \\
\hline
\textbf{Planner} & Requested \textit{Efficacy} [2, 15] and \textit{Implementation} \textbf{[51, 74]}. Instructions focused on CPS readiness and data collection. \\
\hline
\textbf{Retriever} & Fetched [2--15]; reported failure for [51--74]. \\
\hline
\textbf{Summarizer} & \textit{(Truncated)} The law regarding reasonable punishment is confused. Removing the defence clarifies the framework. \\
\hline
\textbf{Critique} & \textcolor{red}{\textbf{Reject:}} Planner requested spans [51-74] which do not exist. Retrieval and Summary are incomplete. \\
\hline
\noalign{\label{sep}} \hline
\multicolumn{2}{|c|}{\textbf{Iteration 2: Refined (Success)}} \\
\hline
\textbf{Planner} & \textbf{Adj:} Scope narrowed to [2, 15]. Focused on \textbf{Public Interest Test} and ``shades of grey." \\
\hline
\textbf{Retriever} & Successfully retrieved turns [5, 9, 13]. \\
\hline
\textbf{Summarizer} & \textit{(Truncated)} Removing the \textit{reasonable punishment} defence aligns law with modern values... avoids a new ``grey area." \\
\hline
\textbf{Critique} & \textcolor{blue}{\textbf{Accept:}} Plan aligned with evidence. Summary captures nuance of legal evolutionary nature and CPS thresholds. \\
\hline
\end{tabular}

\end{table}

\textbf{Error Analysis:} We categorize pipeline failures into a hierarchical taxonomy (Table \ref{tab:taxonomy}) across five domains: Planning (PL), Retrieval (RE), Summarization (SU), Critique (CR), and General Context (CX). This framework enables a granular failure mode effects analysis.

\begin{table}
\centering
\caption{Taxonomy of Failure Modes in the Multi-Agent Pipeline}
\label{tab:taxonomy}
\footnotesize
\begin{tabular}{@{}lll@{}}
\toprule
\textbf{Category} & \textbf{ID} & \textbf{Failure Mode Description} \\ \midrule
Planning (PL) & PL-01 & Span over/under-selection \\
 & PL-02 & Ambiguous or vague instructions \\
 & PL-03 & Neglected critique feedback \\ \midrule
Retrieval (RE) & RE-01 & Missed relevant dialogue turns \\
 & RE-02 & Inclusion of irrelevant content \\ \midrule
Summarization (SU) & SU-01 & Model hallucinations \\
 & SU-02 & Incomplete content coverage \\
 & SU-03 & Missing critical keywords \\ \midrule
Critique (CR) & CR-01 & False acceptance of errors \\
 & CR-02 & Provision of partial feedback \\
 & CR-03 & Non-actionable guidance \\ \midrule
Context (CX) & CX-01 & Drift into unrelated topics \\ \bottomrule
\end{tabular}

\end{table}

\begin{table}[ht]
\caption{Agentic Trace: Iterative Error Identification and Correction. Question: ``Summarise the efficacy of the law.''}
\label{tab:agent_ex1}
\centering
\scriptsize
\renewcommand{\arraystretch}{1.4}
\begin{tabular}{|l|p{5.7cm}|}
\hline
\multicolumn{2}{|c|}{\textbf{Iteration 1 (Failed)}} \\ \hline
\textbf{Planner}    & Requested ``Efficacy'' [2--15] and ``Implementation'' \textbf{[51--74]}. The latter span does not exist in context. \\ \hline
\textbf{Retriever}  & Fetched [2--15]. Reported failure to locate content for span [51--74]. \\ \hline
\textbf{Summarizer} & Summary focuses on efficacy but fails to address implementation instructions. \\ \hline
\textbf{Critique}   & \textbf{Reject:} Flags planner overreach ([51-74]) as the root cause of downstream incompleteness. \\ \hline
\textbf{Error(s)}   & \textbf{PL-01, RE-01, SU-02} \\ \hline
\noalign{\smallskip} \hline
\multicolumn{2}{|c|}{\textbf{Iteration 2 (Success)}} \\ \hline
\textbf{Planner}    & \textit{\textbf{Refined:} Removed span [51--74]. Updated instructions to focus on Public Interest Test nuances.} \\ \hline
\textbf{Retriever}  & \textit{Successfully retrieved efficacy turns [5, 9, 13] with zero noise.} \\ \hline
\textbf{Summarizer} & \textit{Comprehensive summary grounded in legal evolution and CPS thresholds.} \\ \hline
\textbf{Critique}   & \textit{\textbf{Accept:} Confirms alignment between query, plan, and evidence.} \\ \hline
\textbf{Error(s)}   & \textit{None} \\ \hline
\end{tabular}

\end{table}

\begin{table}[ht]
\caption{Agentic Trace: Iterative Performance for Signal Processing Discussion. Question: ``What did the professor think about the Wiener filter?''}
\label{tab:agent_ex3}
\centering
\scriptsize
\renewcommand{\arraystretch}{1.4}
\begin{tabular}{|l|p{5.7cm}|}
\hline
\multicolumn{2}{|c|}{\textbf{Iteration 1 (Failed)}} \\ \hline
\textbf{Planner}    & Selected ``Spectral subtraction'' [402--415]. Instructions focus on Wiener filter, signal subspace, and cascaded system concepts. \\ \hline
\textbf{Retriever}  & Successfully retrieved Professor C’s queries and PhD A’s clarifications on filtering and KL transform. \\ \hline
\textbf{Summarizer} & Reports Wiener filtering and signal subspace relations, but \textbf{omits} the specific conceptualization of the cascaded system. \\ \hline
\textbf{Critique}   & \textbf{Reject:} Validates retrieval quality but identifies the summary as incomplete (SU-02) due to the cascaded system omission. \\ \hline
\textbf{Error(s)}   & \textbf{SU-02} \\ \hline
\noalign{\smallskip} \hline
\multicolumn{2}{|c|}{\textbf{Iteration 2 (Success)}} \\ \hline
\textbf{Planner}    & \textit{\textbf{Refined:} Explicitly mandates the inclusion of the cascaded system (Spectral Subtraction + KL + Wiener).} \\ \hline
\textbf{Retriever}  & \textit{Re-verifies turns [402--415] to ensure all components of the cascaded proposal are captured.} \\ \hline
\textbf{Summarizer} & \textit{Comprehensive summary detailing the Professor's proposal of a cascaded system and its theoretical foundations.} \\ \hline
\textbf{Critique}   & \textit{\textbf{Accept:} Confirms the summary now satisfies all points in the original planning instructions.} \\ \hline
\textbf{Error(s)}   & \textit{None} \\ \hline
\end{tabular}

\end{table}

To illustrate how the proposed error taxonomy interacts with iterative refinement, Table \ref{tab:agent_ex1} presents a step-by-step trace. In Iteration 1, the Planner commits a Span Over-selection (PL-01) error by requesting irrelevant utterances [51–74]. This propagates downstream: the Retriever cannot find the irrelevant content (RE-01), resulting in an incomplete summary (SU-02). Importantly, the Critique agent correctly attributes the root cause to PL-01 rather than blaming the Summarizer, enabling the Planner to refine its instructions in Iteration 2. This highlights the pipeline’s ability to recover from “hallucinated planning” through targeted feedback loops.

Table \ref{tab:agent_ex3} provides a qualitative trace of the \textsc{ReSUM} framework handling a complex technical query regarding signal processing. In the first iteration, although evidence from turns [402–415] was retrieved, the Summarizer missed the “cascaded system” architecture, causing an Incomplete Coverage (SU-02) error. In a standard end-to-end pipeline, this omission would go unnoticed, yielding a technically narrow response. In our framework, the Critique agent detects the mismatch with the Planner’s technical requirements, rejects the draft, and prompts a second iteration where the Planner’s refined instructions explicitly enforce inclusion of the cascaded component.

This iterative process offers two key advantages: (1) Contextual Faithfulness, preserving technical nuances across agent boundaries, and (2) Self-Correction, enabling the system to recover from semantic omissions through modular refinement rather than restarting the entire pipeline.

\section{Conclusion}

In this paper, we have introduced \textsc{\textsc{ReSUM}}, a multi-agent framework for QFS of long, structured meeting transcripts. \textsc{\textsc{ReSUM}} integrates a Planner for task decomposition and traversal planning, a reasoning-guided Retriever with COT distillation for precise evidence selection, a Summarizer for evidence-grounded abstraction, and a Critique agent for iterative evaluation and refinement. By tightly coupling planning, retrieval, summarization, and critique within a structured execution loop, \textsc{\textsc{ReSUM}} produces more faithful, comprehensive, and query-aligned summaries than existing approaches. Experiments on QMSum validate the effectiveness of this reasoning-driven, multi-agent design, while ablation studies highlight the critical role of each agent. Future work will explore extending \textsc{\textsc{ReSUM}} to low-resource conversational settings, reducing reliance on teacher models, and developing lightweight planning and critique mechanisms for efficient real-time deployment.

\section{Limitations}

While \textsc{\textsc{ReSUM}} achieves strong performance on long, structured meeting transcripts, several limitations remain. First, it relies on high-quality teacher LLMs to provide COT rationales for retrieval and summarization, which may limit applicability in low-resource settings or domains where such teachers are unavailable or costly. Second, although multi-agent planning and critique enable iterative refinement, the framework incurs additional computational overhead compared to standard end-to-end summarization models, posing challenges for real-time deployment. Finally, while the Critique agent helps reduce hallucinations and coverage errors, it remains an automated evaluator and may fail to capture subtle semantic inconsistencies or user-specific preferences, limiting full alignment with human judgment.

\section{Generative AI Use Disclosure}
We used ChatGPT as an AI-assisted writing tool to improve language clarity, grammar, and presentation throughout the paper. The model was not used to generate experimental results or analysis, and all technical content, claims, and conclusions were verified by the authors.

\bibliographystyle{IEEEtran}
\bibliography{mybib}

\end{document}